\documentclass[apjl]{emulateapj}
\usepackage{graphicx,subfigure,amsmath, amsfonts, amssymb,aas_macros,times}
\usepackage[usenames,dvipsnames]{color}
\usepackage[export]{adjustbox}
\def\nar{{New A Rev.}}          

\shorttitle{Quenching and the WISE IR Transition Zone}
\shortauthors{K. Alatalo et al.}

\begin{document}


\title{Catching quenching galaxies: The nature of the WISE infrared transition zone}
\author{Katherine Alatalo,$^{1}$ Sabrina~L. Cales,$^{2,3}$ Philip~N. Appleton,$^{1,4}$ Lisa J. Kewley,$^{5}$ Mark Lacy,$^{6}$ Ute Lisenfeld,$^{7}$ Kristina Nyland,$^{8,9}$ \& Jeffrey~A. Rich$^{1,10}$
}

\affil{
$^{1}$Infrared Processing and Analysis Center, California Institute of Technology, Pasadena, California 91125, USA\\
$^{2}$Department of Astronomy, Faculty of Physical and Mathematical Sciences, Universidad de Concepci\'{o}n, Casilla 160-C, Concepci\'{o}n, Chile\\
$^{3}$Yale Center for Astronomy and Astrophysics, Physics Department, Yale University, New Haven, CT 06511 USA\\
$^{4}$NASA Herschel Science Center, California Institute of Technology, Pasadena, California 91125, USA\\
$^{5}$Research School of Astronomy and Astrophysics, Australian National University, Cotter Rd., Weston ACT 2611, Australia\\
$^{6}$National Radio Astronomy Observatory, 520 Edgemont Road, Charlottesville, VA 22903, USA\\
$^{7}$Departamento de F\'isica Te\'orica y del Cosmos, Universidad de Granada, Granada, Spain\\
$^{8}$Physics Department, New Mexico Tech, Socorro, NM 87801, USA\\
$^{9}$National Radio Astronomy Observatory, 1003 Lopezville Road, Socorro, NM 87801, USA\\
$^{10}$Observatories of the Carnegie Institution of Washington, 813 Santa Barbara Street, Pasadena, CA 91101, USA\\
}
\slugcomment{Accepted to the Astrophysical Journal Letters, September 6, 2014}
\email{email:kalatalo@caltech.edu}

\begin{abstract}
We present the discovery of a prominent bifurcation between early-type galaxies and late-type galaxies, in [4.6]--[12] micron colors from the Wide Field Infrared Survey Explorer (WISE).  We then use an emission-line diagnostic comparison sample to explore the nature of objects found both within, and near the edges of, this WISE infrared transition zone (IRTZ).  We hypothesize that this birfurcation might be due to the presence of hot dust and PAH emission features in late-type galaxies.
Using a sample of galaxies selected through the Shocked Poststarburst Galaxy Survey (SPOGS), we are able to identify galaxies with strong Balmer absorption (EW(H$\delta$)$>$5\AA) as well as emission lines inconsistent with star formation (deemed SPOG candidates, or SPOGs*) that lie within the optical green valley.  Seyferts and low ionization nuclear emission line regions, whose $u-r$ colors tend to be red, are strongly represented within the IRTZ, whereas SPOGs* tend to sit near the star-forming edge.  Although AGN are well-represented in the IRTZ, we argue that the dominant IRTZ population are galaxies that are in late stages of transitioning across the optical green valley, shedding the last of their remnant interstellar media.
 
\end{abstract}


\keywords{galaxies: evolution --- galaxies: ISM --- galaxies: star formation --- infrared: galaxies}



\section{Introduction}
The present-day galaxy population has a bimodal color distribution, with a genuine lack of intermediate color galaxies \citep{holmberg59,roberts+94,strateva+01,baldry+04}.  The dearth of green valley objects combined with the increase in passive red galaxies since $z\sim1$ implies that galaxies quench star formation (SF) and rapidly transition from blue to red optical colors \citep{bell+04,faber+07}.  The color bimodality is also associated with a morphological bimodality, with early-type (elliptical) galaxies exhibiting red colors, and late-type (spiral) galaxies exhibiting blue colors.  \citet{schawinski+14} suggest that much of the green valley is a byproduct of this bimodality, with most late-type galaxies transitioning to redder colors slowly as their gas supplies are cut off and a minority of galaxies requiring rapid morphological and color transitions from the blue cloud to the red sequence, making the green valley selection alone incapable of definitively identifying morphologically transitioning galaxies. 

\begin{figure*}[t]
\includegraphics[width=\textwidth,clip,trim=1.1cm 0.8cm 0.7cm 1.1cm]{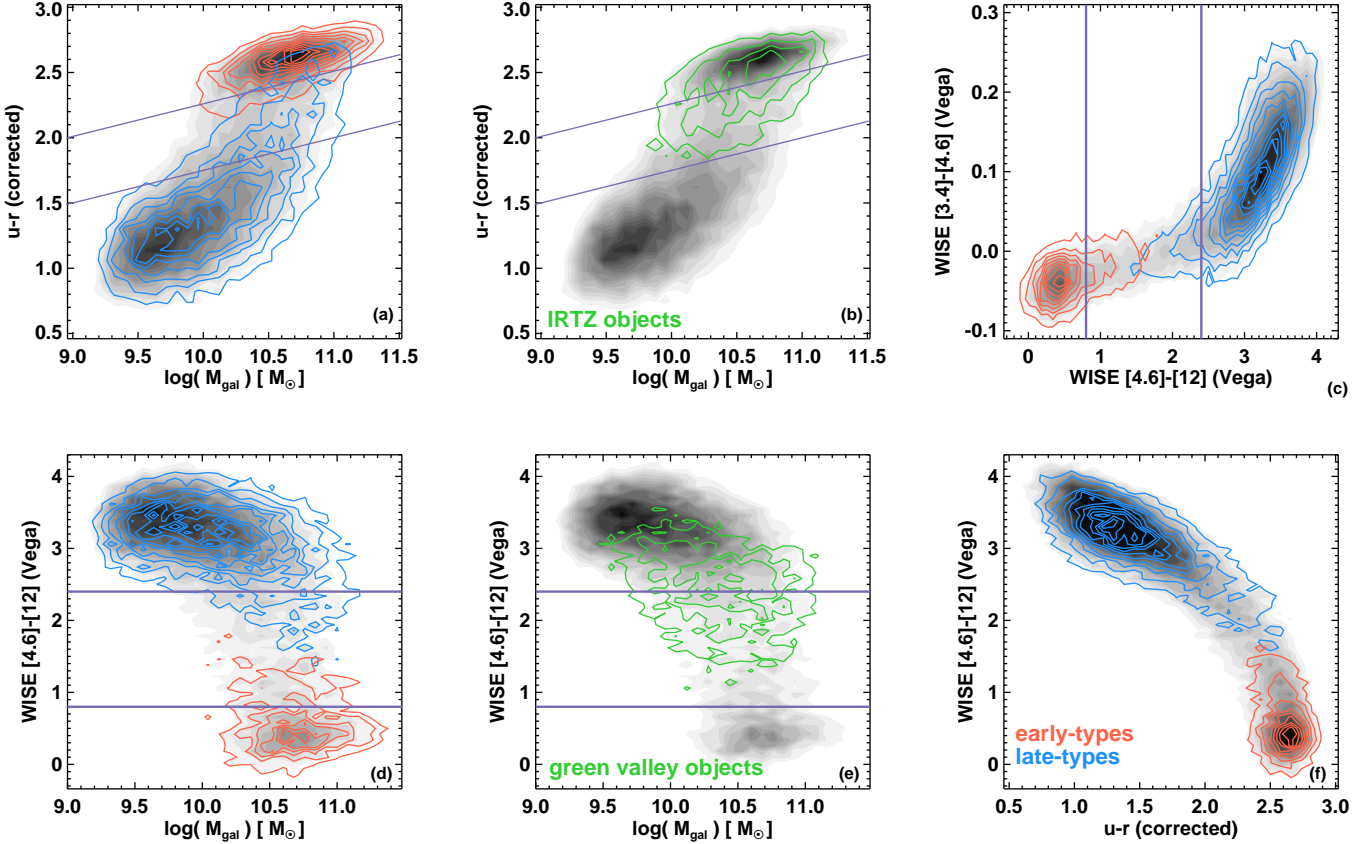}
\caption{The $u-r$ color-mass diagram (a) of WISE-detected Galaxy Zoo galaxies from \citet{schawinski+14}, overlaid with the $u-r$ green valley boundary (purple), reproducing the early-type galaxy/late-type galaxy color bimodality they reported. The same color-mass plot (b) has also been plotted, with contours representing objects classified located in the IRTZ (green contours), showing that IRTZ objects are often green valley objects, though also trend red.
The Galaxy Zoo objects are also plotted in WISE $4.6\mu$m--$12\mu$m vs. $3.4\mu$m--$4.6\mu$m color-color space (c), and show that there is a bimodality in the WISE 4.6$\mu$m and 12$\mu$m colors, with a gap defined to be [4.6]--[12] $\approx$ 0.8--2.4. The [4.6]-[12] color-mass diagram (d) shows a strong bifurcation between early- and late-type galaxies, with a tail of late-types that leaks into the transition zone, though the overlap that is seen in the $u-r$ colors is much less pronounced.  The same [4.6]-[12] color-mass plot is reproduced (e) , with green contours representing objects that were classified as green valley objects.  In this case, optically classed green valley objects tend to fall closer to the star-forming segment of the [4.6]-[12] color plot.  The $u-r$ vs. [4.6]--[12] color-color plot (f) shows that there is a tight relation between the colors, and that the transition zone is visible as the elbow in the relation. Contours are at 5\% intervals of the maximum of the total distribution, except in panels (b) and (e), which have 20\% contours.}
\label{fig:wisecmd}
\end{figure*}

Poststarburst galaxies (i.e. K+A and E+A; \citealt{dressler+83,zabludoff+96,quintero+04,goto05}), the archetypical green valley population, have been identified as morphological transition objects (EW(H$\delta) > 5$\AA, and a lack of [O~{\sc ii}] and/or H$\alpha$ emission lines), but these searches select against ionized gas emission lines.  Recently, the traditional definition for poststarburst galaxies has been found to be too narrow to encompass the full range of poststarburst transitioning objects \citep{falkenberg+09}.  \citet{cales+11} were able to show that while quasars would be rejected from traditional poststarburst searches due to the presence of [O~{\sc ii}] emission, they are indeed a part of the transitioning population of galaxies.  \citet{alatalo+11,alatalo+14} show that Active Galactic Nucleus (AGN) outflow host NGC~1266 is also a galaxy transitioning from a star-forming late-type galaxy to an early-type galaxy, and has low ionization nuclear emission line region (LINER)-like emission due to shocks \citep{davis+12}.  This galaxy would have been missed by traditional poststarburst searches.

\begin{figure*}[t!]
\includegraphics[width=\textwidth]{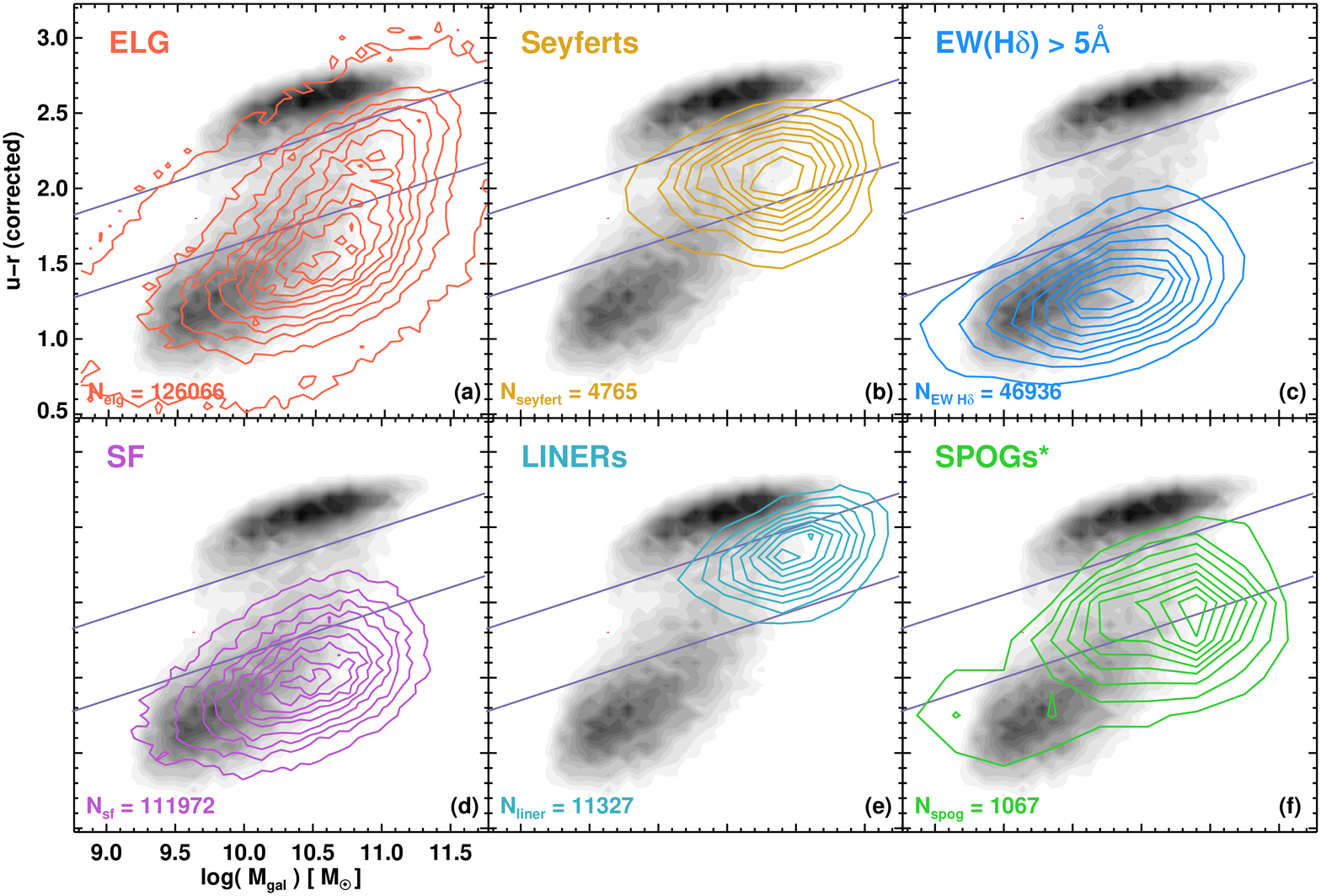}
\caption{The corrected $u-r$ color-mass diagram calculated from the universal sample of early- and late-type galaxies from \citet{schawinski+14} (grayscale) overlaid with the ELG sample (a: red), Seyferts (b: yellow), H$\delta$ absorbers (c: blue), SF (d: violet), LINERs (e: turquoise) and SPOGs* (f: green). Contours for the underlying Galaxy Zoo sample are increments of 5\% of the maximum bin.  Contours for all other subsamples are increments of 10\% of the maximum, though the lowest contour on the ELG relation starts at 1\%.  The mass and $u-r$ colors were $k$-corrected and corrected for extinction (see text).  Galaxies with star formation (SF or H$\delta$ objects) can be distinguished as part of the blue cloud, LINERs appear to be most concentrated in the red sequence, {\bf though peak in the green valley}, while Seyfert objects are found in the green valley.  SPOGs* are seen on the blue edge of the green valley.  Numbers in the bottom left corner of each panel represent the total number of subsample galaxies with robust detections in $u$, $r$ and $K_s$.  The green valley is overplotted in indigo.}
\label{fig:spogscmd}
\end{figure*}

{\em Spitzer} enabled a detailed study of galaxies in the infrared (IR), and \citet{lacy+04} and \citet{stern+05} showed that an {\em Spitzer-}based IR color diagnostic was able to identify the presence of an AGN, though there was not an obvious bimodality seen in the original {\em Spitzer} work between early- and late-type galaxies. \citet{assef+13} showed that enhanced WISE colors ($[3.4]-[4.6]>0.8$) could definitively identify an AGN, but this was only sensitive to the brightest AGNs.  An IR gap was first noted in Hickson Compact Groups \citep{johnson+07,walker+13}.  \citet{cluver+13} discovered that galaxies with warm H$_2$ excesses (as measured by the H$_2$/PAH flux from the mid-IR) tended to fall in the IR gap, hypothesizing that shocks heated the H$_2$.  Despite this early evidence that transitions might be occurring in group galaxies, detailed studies searching for a gap in IR colors of morphologically classified galaxies has not been reported.
 
 


\section{The Sample}
\label{sample}
The comprehensive early- and late-type galaxy samples compiled by \citet{schawinski+14} from the Galaxy Zoo (\citealt{lintott+08,lintott+11}) were used as the underlying comparison sample.  This sample demonstrated how different morphological types manifest in IR color space, and was used to compare to the selected sub-samples described below.  The sample included the corrected $u-r$ colors and the stellar masses of 47,995 early- and late-type galaxies.  The Galaxy Zoo sample was then cross-correlated with the WISE all-sky catalog \citep{wise}, which successfully matched 47,729 (99\%).  A 5\arcsec\ matching radius was used for the \citet{schawinski+14} sample, given its local ($z=0.02-0.05$) redshift range, and the likelihood that many of the galaxies are extended in both WISE and SDSS.  In both the following work and \citet{schawinski+14}, we use cosmological parameters $H_0 = 70$~km~s$^{-1}$, $\Omega_m = 0.3$ and $\Lambda = 0.7$ \citep{wmap}.

\begin{table}[b]
\caption{Detections Rates in WISE}
\centering
\begin{tabular}{r| c c c c | c}
\hline \hline
& W1 & W2 & W3 & All & W1--W2\\
(\%) & 3.4$\mu$m & 4.6$\mu$m & 12$\mu$m & 3 &$>$0.8$^\dagger$\\
\hline
Galaxy Zoo & 99 & 99 & 85 & 85 & 0.1\\
Late-types & 99 & 99 & 98 & 98 & 0.1\\
Early-types & 99 & 99 & 70 & 70 & 0.1\\
\hline
ELG & 87 & 87 & 82 & 82 & 0.2\\
Seyferts & 91 & 91 & 85 & 85 & 3.4\\
EW(H$\delta$)$>$5\AA & 89 & 89 & 83 & 83 & 0.1\\
SF & 87 & 87 & 84 & 84 & 0.1\\
LINERs & 92 & 92 & 71 & 71 & 0.1\\
SPOGs* & 89 & 89 & 80 & 80 & 2.1\\
\hline \hline
\end{tabular}
\raggedright
\noindent$^\dagger$Mid-infrared selected AGN criterion based on \citet{assef+13}
\label{tab:wise_det}
\end{table}

\begin{figure*}[t]
\includegraphics[width=\textwidth]{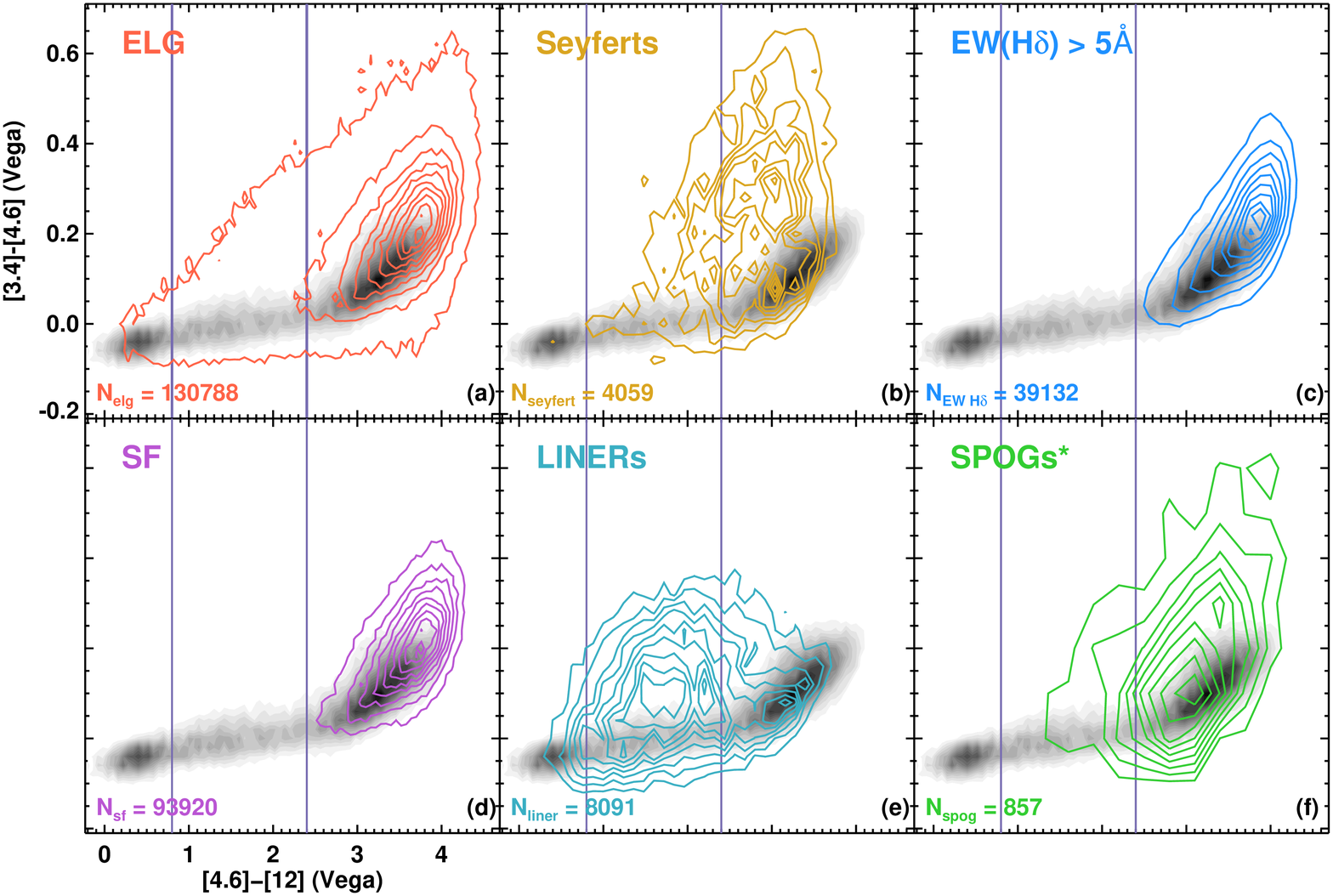}
\caption{[4.6]--[12] vs. [3.4]--[4.6] WISE color-color diagram of the Galaxy Zoo sample (grayscale), the ELG sample (a: red), Seyferts (b: yellow), H$\delta$ absorption (c: blue), star formation (d: violet), LINERS (e: turquoise) and SPOGs* (f: green).  Contours for the Galaxy Zoo sample are increments of 5\% of the maximum bin.  Contours for all other subsamples are increments of 10\% of the maximum, though the ELG panel includes a 1\% contour to show the approximate total range of objects.  Numbers in the bottom left corner of each panel represent the total number of subsample galaxies with 3.4, 4.6 and 12$\mu$m SNR$>3$. The IRTZ is overplotted in indigo.}
\label{fig:spogswise}
\end{figure*}

For the spectrally-selected sample we used the Oh-Sarzi-Schawinski-Yi (OSSY) catalog \citep{ossy}\footnote{http://gem.yonsei.ac.kr/$\sim$ksoh/wordpress/}, which spans redshifts between $z=0.02-0.2$, and restricted ourselves only to objects detected with a signal-to-noise ratio $>$3 in all diagnostic lines (H$\alpha$, H$\beta$, [O~{\sc iii}], [O~{\sc i}], [N~{\sc ii}], and [S~{\sc ii}]), hereafter known as the Emission Line Galaxy (ELG) sample).  A total of 159,387 (24\% of the OSSY sample; Cales et al. 2014, in preparation ) galaxies matched these criteria.  The subsamples were then classified based on the models of \citet{kewley+06}, including SF, LINERs, and Seyferts, based on their emission line ratios.  An additional subsample was also defined, based on Balmer absorption properties (EW(H$\delta)>5$\AA).  The fifth subsample selected to detect transitioning galaxies is described below.

We started the search for transitioning objects through the Shocked POststarburst Galaxy Survey (SPOGS; Cales et al. 2014, in preparation), which identifies galaxies within OSSY \citep{ossy} that might contain both shocked gas and a poststarburst stellar population, a combination present NGC~1266 \citep{davis+12,alatalo+14}.  SPOGS selects objects based on the presence of Balmer absorption (EW(H$\delta)>5$\AA; indicative of an A-star population; \citealt{goto05,tremonti+07}) and emission line ratios \citep{bpt,veilleux+87} consistent with shocks \citep{allen+08,rich+11}. SPOGS currently excludes galaxies with emission line diagnostics consistent with either SF-only or composite SF+AGN models, where contaminants dominate the distribution.

Of the 159,387 ELG galaxies, 1,067 were classified as SPOG candidates, as the SPOGS criteria create a subsample that is not necessarily shocked.  In this paper, we refer to these candidates as SPOGs*, since ongoing observations are necessary to confirm whether the excitation of the ionized gas is indeed due to shocks (Cales et al. 2014, in preparation).

The ELG sample (and subsamples) were then cross-correlated with the WISE catalog, with match percentages listed in Table \ref{tab:wise_det}.  The Galaxy Zoo sample and the majority of the ELG sample have detection rates with WISE of at least 85\%.  According to the WISE catalog, 72\% of the Galaxy Zoo, and 28\% if the ELG catalogs, are extended in at least one WISE band.  Since extended source fluxes can be misrepresented by the {\em w$\star$mpro} value \citep{cluver+14}, we use the fluxes derived from fitting ellipses ({\em w$\star$gmag}, when available) based on the 2MASS source. 

All $u$ and $r$ magnitudes, and $u-r$ colors for the ELG sample were {\em k}--corrected with the IDL routine {\tt calc\_kcor.pro}\footnote{http://kcor.sai.msu.ru/} \citep{calc_kcor}.  We then corrected for extinction using the spectral fit-derived E(B-V)$_{\rm stars}$ output by the OSSY catalog \citep{ossy}, and converted to extinction at the rest wavelength of each object using the relation in \citet{calzetti+00}.

The early-type subset of the Galaxy Zoo sample and the LINER ELG subsample contained the lowest W3/12$\mu$m detection rate, with $\approx$70\% detections with signal-to-noise $>3$.  Despite early-types typically being ISM-poor (and therefore faint in the mid-IR), WISE has detected a majority of these systems at 12$\mu$m.  


\section{Results and Discussion}
\label{resdisc}
\subsection{Galaxy Type Bimodality in WISE [4-6]-[12] Colors}
\citet{wise} provided a schematic of different astrophysical objects within the [3-4]--[4.6]  vs. [4.6]--[12] WISE color space, with late-type [4.6]--[12] color between $1<[4.6]-[12]<4$ magnitudes and early-type [4.6]--[12] colors clustered tightly around 1. Figure \ref{fig:wisecmd} uses the classifications of the Galaxy Zoo sample to show that, while there is an overlap between the populations in $u-r$ colors, there is very little overlap between the late- and early-type galaxy WISE colors, with a [4.6]--[12] color gap between {\bf $\approx 0.8-2.4$}, demarcated in Fig. \ref{fig:wisecmd}b \& c as indigo lines, and deemed an ``infrared transition zone'' (IRTZ).  

At least at low redshift, a WISE color selection seems able to differentiate between galaxy populations and identify objects within the IRTZ.  While most objects avoid this region, the tails of the galaxy distributions which do begin to inhabit the IRTZ are either low mass early-type galaxies or high mass late-type galaxies, and it appears that there is very little overlap, unlike the $u-r$ optical colors. 

Figure \ref{fig:wisecmd}f includes the $u-r$ vs. [4.6]--[12] color diagram, which shows that there is a clear relation between colors.  In fact, it appears that although there is a large spread in $u-r$ colors in late-type galaxies and a larger spread in [4.6]--[12] colors in early-type galaxies, the transition zone in both colors is clearly identifiable.  

\begin{figure*}
\includegraphics[width=\textwidth]{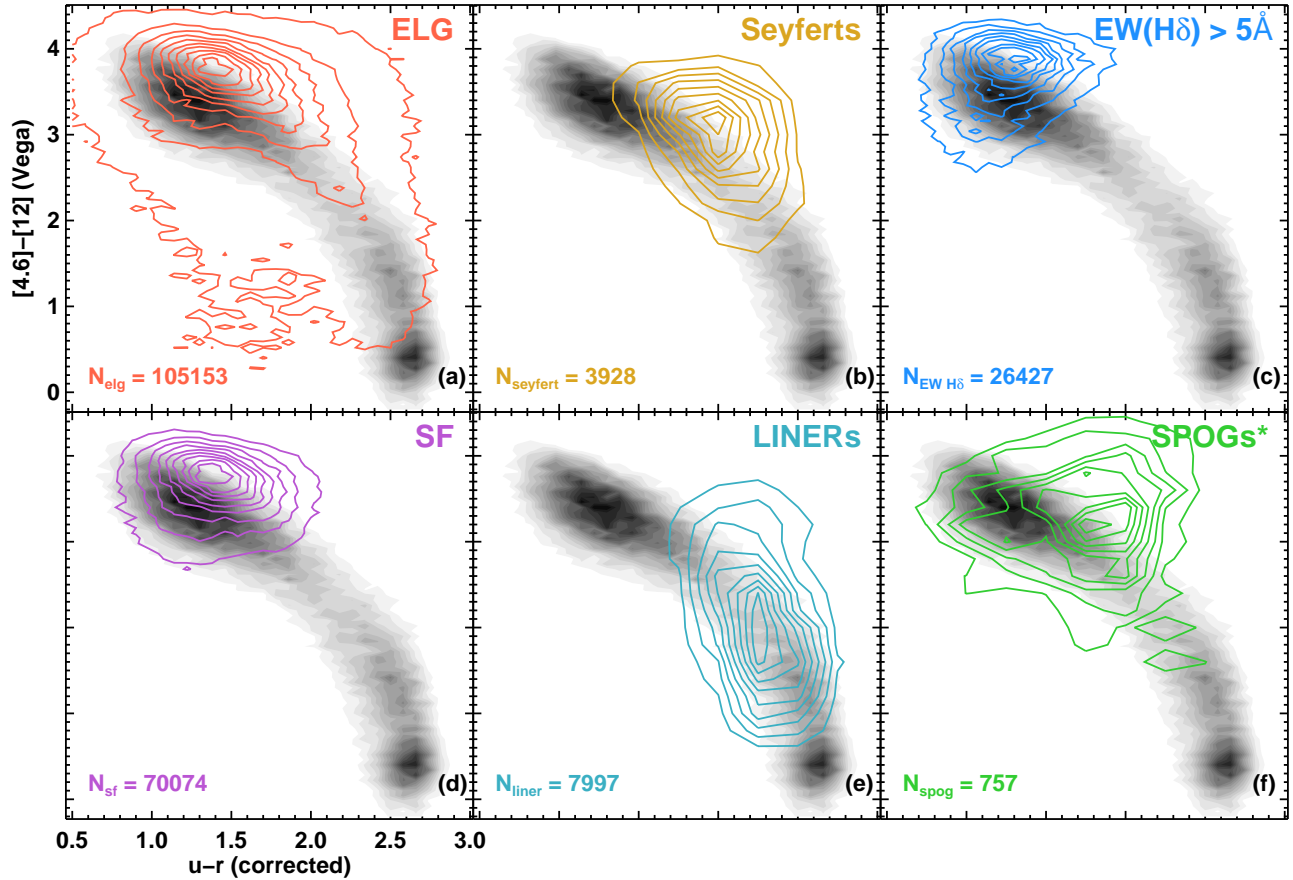}
\caption{W2--W3 vs. $u-r$ color-color diagram from Galaxy Zoo (grayscale), the ELG sample (a: red), Seyferts (b: yellow), the Balmer absorbers (EW(H$\delta>5$\AA; blue, c), SF (d: violet), LINERs (e: turquoise) and SPOGs* (f: green).  Contours for the Galaxy Zoo sample are increments of 5\% of the maximum bin.  Contours for all other subsamples are increments of 10\% of the maximum, with a 1\% contour on the ELG sample.  Numbers in the bottom left corner of each panel represent the total number of subsample galaxies with 3.4, 4.6 and 12$\mu$m SNR$>3$.  The $u-r$ colors relate well to the [4.6]--[12] colors, and SF and EW(H$\delta)>5$\AA\ clearly lie in the SF portion of the color-color plot, with Seyferts populating the transition zone in both, and LINERs showing optically red colors and transitioning IR colors.  SPOGs* also have transitioning colors, though have a much larger spread.}
\label{fig:spogscolors}
\end{figure*}

It is likely that this IRTZ comes from a combination of properties, including the presence of the PAH features (7.7, 8.6, 11.3 and 12.7$\mu$m are in the W3 channel), a non-negligible hot dust contribution at 12$\mu$m (whereas the emission at 4.6$\mu$m is still strongly dominated by stars), and possibly the contribution of an AGN to the IR colors, though the lack of coincident changes in the [3.4]--[4.6] colors (Fig. \ref{fig:wisecmd}c; \citealt{donley+12}) seems to indicate that AGNs are not the dominant contribution.  On a larger scale, this result simply implies that early-type galaxies tend to be ISM poor, and late-type galaxies tends to be ISM rich, and that the action of losing observational signatures of the ISM tend to be rapid.  The fact that the tail end of the distribution of massive late type galaxies resides in the IRTZ suggests that these objects are currently quenching SF.

\subsection{SPOGs* and the Optical Green Valley}

Figure \ref{fig:spogscmd} shows the distributions across $u-r$ color-mass space with the Galaxy Zoo sample (grayscale; \citealt{schawinski+14}). Given that the ELG sample spans higher redshift and requires non-negligible detections of ionized gas in all lines, the fact that it skews to massive, blue galaxies compared to the Galaxy Zoo sample is unsurprising.

EW(H$\delta) > 5$ \AA\ galaxies require a large contribution from A-stars, creating a significant overlap between these galaxies and those classed as purely SF.  This means that most Balmer absorbers are star-formers.  LINERs in the ELG sample peak toward the red side of the green valley, which could be due to the stringent emission line cutoff we have employed, biasing us toward LINERs with large ionized gas fluxes, and potentially selecting star formation+AGN composites \citep{ho2008} or shocks. While the majority of LINERs in this sample are likely early-type galaxies with either low luminosity AGNs \citep{kewley+06} or post-asymptotic giant branch star contributions \citep{yan+06}, the fact that the LINERs in the ELG sample are also greener on-average might be evidence that many also exhibit shocks \citep{heckman80}.  Seyferts within the ELG sample appear to be massive, with a large presence in the green valley.  Given that moderate AGN activity has long been known to correlate with galaxies with transitional colors \citep{heckman+06,schawinski+09}, the location of the Seyferts is unsurprising.

Figure \ref{fig:spogscmd}f shows that SPOGs* peak at the blue boundary of the green valley, and in a completely different location from the star-forming objects, supporting the idea that this new selection has identified transitioning objects, although the excitation mechanism for the ionized gas in SPOGs* might not predominantly be shocks.  Seyfert contamination could be an issue, though only 18\% of SPOGs* have pure Seyfert line diagnostics.  Thus, the majority of SPOGs* are not simply Seyferts with poststarburst stellar populations.  Future observations focusing on the near-IR H$_2$ emission (similar to what was observed in NGC~1275; \citealt{hatch+05}) will be able to differentiate between shocks and other excitation mechanisms, giving us the ability to fully understand the SPOG* population.


\subsection{The constituents of the Infrared Transition Zone (IRTZ)}
Figure \ref{fig:spogswise} shows the WISE IR colors of each subsample, indicating a variety of distributions of IR colors depending on emission line diagnostics.  The ELG sample spans the range of WISE IR colors, though the distribution is predominantly consistent with the late-type galaxies (from Fig. \ref{fig:wisecmd}c), as well as the SF and EW(H$\delta)>5$\AA\ sources, with a tight locus of points around $[3.4]-[4.6]\approx0.25$ and $[4.6]-[12]\approx3.8$, as well as blue $u-r$ colors. 

\citet{donley+12} have argued that AGN spectra show a flat power law in the IR, and Fig.~\ref{fig:spogswise}b shows that Seyferts appear consistent with this interpretation.  While a pure WISE color diagnostic is only able to identify the brightest AGNs, the spread in $[3.4]-[4.6]$ and [4.6]--[12] colors of the Seyferts (compared to late- and early-type galaxies) is consistent with the contribution of an AGN to the IR emission of a galaxy.  Seyferts also tend to be found near the edge of the IRTZ, likely from a combination of the AGN IR contribution (shifting both $[3.4]-[4.6]$ and $[4.6]-[12]$ seen in Fig. \ref{fig:spogswise}b; \citealt{donley+12}) and transitioning colors (also evident in $u-r$ from Fig. \ref{fig:spogscmd}b).

WISE colors of the LINERs also show a substantial spread in WISE color space, but are more similar to early-types than other objects in the ELG sample.  \citet{sturm+06} and \citet{kaneda+08} have shown that many LINERs show strong 11.3$\mu$m PAH features, which need only a small UV field  to excite, shifting its color toward the SF cloud.  LINER emission from low-luminosity AGNs could also contribute to the $[3.4]-[4.6]$ and $[4.6]-[12]$ colors in many cases, shifting those colors redward.  \citet{sturm+06} also found that in many cases the LINERs found to be IR-faint also showed supernova remnant-like IR line diagnostics, which are thought to originate from shocks.  LINER emission from shocks could also be present in this subsample \citep{cluver+13}.

SPOGs* appear to be primarily located close to the edge of the IRTZ than SF-dominant systems (moreso than LINERs and Seyferts).  This could be in part due to the presence of an AGN, inferred based on the larger spread in [3.4]--[4.6] colors, and in fact, NGC~1266, the prototypical SPOG has an influential AGN \citep{alatalo+11}.  SPOGS* have green $u-r$ colors, and SF-like [4.6]--[12] colors, but sit apart from the SF galaxies.  This $u-r$ [4.6]--[12] color combination might imply that SPOGs* stellar populations are beginning to transition, but could also be influenced by the presence of an AGN.

\subsection{The nature of the IRTZ}

Figure \ref{fig:spogscolors} shows the $u-r$ vs [4.6]--[12] colors with the different line diagnostic classifications overlaid upon the Galaxy Zoo sample.  As in Fig. \ref{fig:wisecmd}f, the correlation between the colors is tight, and in most cases, the different line classes fall faithfully on the color-color relation, with the SF and Balmer absorbed populations sitting prominently in the SF color cloud.  Seyferts sit in the color transition zone, without much scatter out of the relation defined by the Galaxy Zoo sample.  It is possible that these emission detected LINERs represent a population of galaxies that are completing their optical color transitions, but have not yet completely lost their ISMs, which could be observed as an enhancement in their 12$\mu$m emission, while Seyferts and SPOGs* are in earlier stages of the transition.

The nature of the IRTZ is not yet understood, but we suggest some factors that might create both the tight color-relation between IR and optical as well as the IRTZ.  The location of the LINERs and the Seyferts seems to indicate that an AGN is able to influence the IR color of the host galaxy, migrating all IR colors redward, but the fraction of AGNs capable of significantly changing their host's IR colors is small \citep{lacy+04,stern+05,assef+13}. The optical colors of objects other than quasars also should not be dominantly affected by the AGN \citep{ho2008}.  The tight optical-IR relation could also be the result of evolution.  In fact, Figures \ref{fig:wisecmd}b and \ref{fig:wisecmd}e suggest that this is the case, with optical green valley objects lying nearer to the SF portion of the [4.6]--[12] color-mass distribution and IRTZ objects lying closer to the optical red sequence.  This suggests that galaxies transition across optical color space as the SFR begins to diminish, and later completely shed their ISMs (likely the dominant constituent of the 12$\mu$m emission), since 12$\mu$m can be enhanced with a minimal incident radiation field and a modicum of the remaining ISM.  More detailed studies of the stellar population ages of galaxies within the IRTZ are planned, but beyond the scope of this paper.

\section{Summary}
The presence of both a tight relation between $u-r$ colors and [4.6]--[12] colors, as well as a prominent transition zone, visible in IR and definitive in $u-r$ vs [4.6]--[12], indicates that the addition of WISE colors to classifications of galaxies is able to identify transitioning sources through a new classification scheme.  Spectrally classed objects, for the most part, fall into the regions of the color space in expected ways.  The position of the transitioning objects in each plot also seem to point to an evolutionary picture, where galaxies transition first in optical colors and then in IR colors.  The SPOGs*, which are selected to be transitioning objects follow this trend as well, with green $u-r$ colors at the boundary of the green valley, and on the edge of the SF cloud, very near the IRTZ.
 



\acknowledgments K.A. would like to thank Kevin Schawinski for giving full access to his early \& late-type galaxy sample, as well as Theodoros Bitsakis and Patrick Ogle for insightful discussions, Michelle Cluver and Roc Cutri for WISE advice, and finally the referee for an informative and insightful report that improved the paper.  K.A. is supported by funding through Herschel, a European Space Agency Cornerstone Mission with significant participation by NASA, through an award issued by JPL/Caltech.  K.N. is supported by NSF grant 1109803.  S.L.C. was supported by ALMA-CONICYT program 31110020. U.L. acknowledges  support by the research projects AYA2011-24728 from the Spanish Ministerio de Ciencia y Educaci\'on and the Junta de Andaluc\'\i a (Spain) grants FQM108.  This publication makes use of data products from the Wide-field Infrared Survey Explorer, which is a joint project of the University of California, Los Angeles, and the Jet Propulsion Laboratory/California Institute of Technology, funded by the National Aeronautics and Space Administration.  This research has made use of the Sloan Digital Sky Survey\footnote{http://www.sdss.org/}.



 \end{document}